\documentclass[fleqn,referee]{mnras}

\usepackage{txfonts}

\usepackage[T1]{fontenc}
\usepackage{ae,aecompl}

\usepackage{graphicx}	
\usepackage{amssymb}	

\title[GX 1+4]{Comprehensive Timing and X-ray Spectral Analysis of GX 1+4}

\author[M. M. Serim, \c{S}. \c{S}ahiner, D. \c{C}erri--Serim, S. \c{C}. \.{I}nam, and A. Baykal ]
{M. M. Serim$^{1}$\thanks{E-mail: muhammed@astroa.physics.metu.edu.tr (MMS); seyda@astroa.physics.metu.edu.tr (\c{S}\c{S}); 
danjela@astroa.physics.metu.edu.tr (D\c{C}S); inam@baskent.edu.tr (S\c{C}\.{I}); altan@astroa.physics.metu.edu.tr (AB)}, 
\c{S}. \c{S}ahiner$^{1}$\footnotemark[1], D. \c{C}erri--Serim$^{1}$\footnotemark[1], S. \c{C}. \.{I}nam$^{2}$\footnotemark[1] 
and A. Baykal$^{1}$\footnotemark[1] \\
$^{1}$Physics Department, Middle East Technical University, 06531 Ankara, Turkey\\
$^{2}$Department of Electrical and Electronics Engineering, Ba\c{s}kent University, 06790 Ankara, Turkey}

\date{Accepted XXX. Received YYY; in original form ZZZ}

\pubyear{2017}

\begin{document}
\label{firstpage}
\pagerange{\pageref{firstpage}--\pageref{lastpage}}
\maketitle

\begin{abstract}

We present analysis of RXTE--PCA observations of GX 1+4 between March 3, 2001 and January 31, 2003 together with the CGRO--BATSE X-ray flux and frequency derivative time series between 1991 and 1999. From the timing analysis of RXTE-PCA observations, we are able to phase connect pulse arrival times of the source within two different time intervals and obtain corresponding timing solutions. Using these pulse arrival times, we contribute to long term pulse frequency history of the source. We look for episodic correlations and anti-correlations between torque and X-ray luminosity using CGRO--BATSE X-ray flux and frequency derivative time series and find that correlation state of GX 1+4 seems to change on $\sim$ 100-200 days long intervals. We estimate torque noise of the source and observe flickering noise ($f^{-1}$). We achieve to measure the longest observed timescale for a noise process among accretion powered X-ray pulsars by extending the noise estimate for a time scale ranging from 31 days to 44 years. Spectral analysis of individual RXTE-PCA observations indicates a significant correlation between iron line flux and unabsorbed X-ray flux. Pulse phase resolved spectra of the source indicate a broadening of iron line complex at the bin corresponding to the pulse minimum. 

\end{abstract}




\section{Introduction}

Accretion powered pulsar GX 1+4 was discovered in 1970 with the pulsations of about two minutes period (Lewin et al. 1971). It showed strong spin-up during 1970s (White et al. 1983). Subsequent to an undetectable low luminosity state era during early 1980s, the source was redetected after it had undergone a torque reversal (Makishima et al. 1988). After this torque reversal event, GX 1+4 has been usually observed to be spinning down (Gonzales-Galan et al. 2012).   

GX 1+4 is in a low mass X-ray binary (LMXB) system and its optical counterpart is an M6III type red giant star V2116 Oph which underfills its Roche Lobe (Glass \& Feast 1973; Chakrabarty \& Roche 1997a; Hinkle et al. 2006). Distance to this system was estimated to be about 4.3 kpc (Hinkle et al. 2006).  

GX 1+4 is the first example of an accretion powered pulsar residing in a symbiotic X-ray binary system in which the compact object accretes mass via the dense wind of the M-type giant companion (Corbet et al. 2008). Presence of a long-term accretion disc around the pulsar was suggested and standard accretion disc theory (Ghosh \& Lamb 1979; Wang 1987) was used to explain the pulse period evolution, torque reversal and transition to faint state (Dotani et al. 1989; Cui \& Smith 2004). 

Magnetic field strength of GX 1+4 was inferred from two methods:  Implementing standard accretion disc theory and using marginal evidence of cyclotron resonance scattering feature (CRSF). With the standard accretion disc theory, the surface magnetic field strength of the pulsar was estimated to be quite high, $\sim 10^{13} - 10^{14}$ Gauss (Dotani et al. 1989; Cui \& Smith 2004). On the other hand using the relation between magnetic field and cyclotron line energy, magnetic field strength was calculated to be rather ordinary among accretion powered pulsars, $\sim 10^{12}$ Gauss (Rea et al. 2005; Ferrigno et al. 2007).  

Although the optical companion of GX 1+4 is known, orbital parameters of the system have not been conclusively determined yet. From X-ray measurements of the spin period variations during both spin up and down era (between 1991 and 1998), an orbital period of 304 days has been suggested (Cutler et al. 1986, Pereira et al. 1999, Braga et al. 2000). On the other hand, infrared observations of the source have indicated a 1161-day period without any sign of 304-day periodicity (Hinkle et al. 2006). Recently, 1161-day period has been supported by suggesting a potential neutron star eclipse from the variability of hard (>17 keV) X-ray emission (Ilkiewicz et al. 2017).

X-ray spectrum of GX 1+4 was studied extensively before. Both a power law with an exponential cutoff (Cui 1997, Kotani et al. 1999, Cui \& Smith 2004, Ferrigno et al. 2007) and a model based on Comptonization of soft photons in hot plasma (Galloway 2000, Galloway et al. 2000, Galloway et al. 2001, Naik et al. 2005) are found to fit the X-ray spectra well. The continuum is influenced by highly variable intrinsic absorption and strong iron line emission. GX 1+4 exhibits low luminosity episodes with irregular time coverage lasting from about one day (Galloway et al. 2000; Rea et al. 2005; Naik et al. 2005) to several months (Cui 1997; Cui \& Smith 2004). During these episodes, iron emission is enhanced and the continuum is generally harder without an indication of cutoff. A BeppoSAX observation of GX 1+4 during a low luminosity state revealed a strong iron K$\beta$ emission along with iron K$\alpha$ (Rea et al. 2005) and a narrower iron K$\beta$ line is found to be present for moderate luminosities as well (Naik et al. 2005).

In this paper, we present timing and spectral analysis of 90 pointing RXTE--PCA observations of GX 1+4 between March 3, 2001 and January 31, 2003 with a total exposure of 276 ks. We also look for episodic correlations and anti-correlations between torque and X-ray luminosity using CGRO--BATSE X-ray flux and frequency derivative time series between 1991 and 1999. In Section
\ref{sect:obs}, we briefly introduce the observations. In Section \ref{subsect:ptiming}, we present pulse timing analysis of RXTE--PCA observations and resulting measurements of pulse frequencies. In Section \ref{subsect:noise}, we present power spectrum of the pulse frequency derivative fluctuations. In Section \ref{subsect:episodic}, we present analysis of CGRO--BATSE data for episodic torque luminosity correlations and anti-correlations. In Sections \ref{subsect:trspectra} and \ref{subsect:pprspectra}, we present analysis of time resolved and pulse phase resolved spectra of RXTE--PCA data. In Section \ref{sect:discuss}, we summarize and discuss our results.

\section{Observations}
\label{sect:obs}

\subsection{RXTE Observations}
  
The \textit{Rossi X-ray Timing Explorer} (RXTE) was an X-ray satellite which had been launched into low-Earth orbit on December 30, 1995 and operated until January 5, 2012. 
It had Proportional Counter Array (PCA) composed of proportional counter units (PCUs), each of which had an effective area of 1300 cm$^2$. The PCA was sensitive to the 
photons within 2--60 keV energy range. GX 1+4 was monitored with RXTE--PCA between March 3, 2001 and January 31, 2003. During this time interval, 90 pointing observations were 
carried out. Total exposure of these observations is around 276 ks while the exposure for each observation varies between 0.5 and 18 ks (see Table \ref{tab:obslog} for details).

\begin{table}
	\centering
	\caption{Observation log of GX 1+4}
	\label{tab:obslog}
	\begin{tabular}{lccc} 
		\hline
		Proposal & Number of  & Total & Time Range \\
		 ID  & observations & exposure (ks) & (MJD)\\
		\hline
		60060 & 40 & 120.9 & 51974-52320\\
		70064 & 40 & 124.6 & 52338-52593\\
		70065 & 8 & 28.9 & 52390-52585\\
		70425 & 2 & 1.5 & 52662-52670\\
		\hline
		total & 90 & 276.0 & 51974-52670\\
	\end{tabular}
\end{table}

The data reduction is carried out with the \texttt{HEASOFT v6.19} software. We select the data considering the electron contamination to be less than 0.1, offset angle to be 
less than 0.02$^\circ$ and elevation angle to be greater than 10$^\circ$. We extract lightcurves from \texttt{GoodXenon} mode events with 1s bin time. Then, barrycentric correction is applied to
the photon arrival times in the lightcurve. As the active number of PCUs in each observation varies, we correct the count rates with \texttt{CORRECTLC} command as if all five 
PCUs were operating at the same time. For the spectral analysis we use \texttt{Standard2f} mode data which have 128 energy channels. Spectra are extracted only from PCU2 data. 
Furthermore, we construct pulse phase resolved spectra with the tool \texttt{FASEBIN}. The PCA  background  estimator  models (Epoch 5a and 5b)  supplied  by  the RXTE Guest Observer 
Facility (GOF) are used for background subtraction. Spectral analysis is performed with \texttt{XSPEC V.12.9.0}. A systematic error of 0.5\% is applied as it is recommended by the PCA team.

\begin{figure*}
	\includegraphics[width=0.4\columnwidth, angle=270]{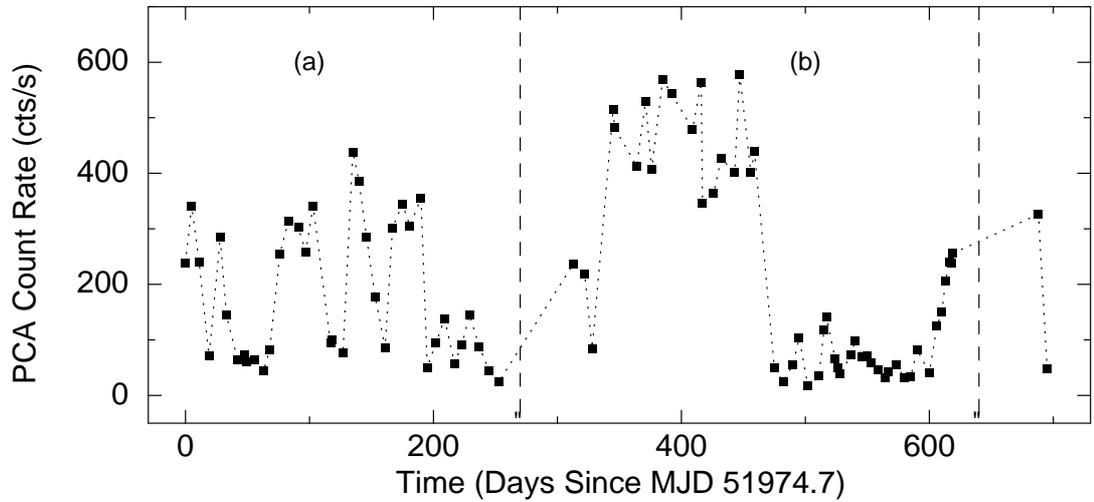}
    \caption{1-day rebinned RXTE--PCA lightcurve of GX 1+4. Dashed lines indicate two different timing ephemeris; interval "a" and interval "b".}
    \label{fig:example_lc}
\end{figure*}

\subsection{CGRO-BATSE}

BATSE (Burst and Transient Source Experiment) onboard CGRO (Compton Gamma Ray Observatory) consisted of eight detector modules located at the corners of CGRO. These detectors enabled continuous all sky monitoring of both
pulsed and unpulsed sources above 20 keV between 1991 and 2000. BATSE
not only continously monitored daily pulse frequency and X-ray flux changes of persistent and previously known transient accretion powered pulsars,    but also discovered and monitored new transients (Bildsten
et al. 1997).

This paper makes use of BATSE 20-60 keV band X-ray flux and pulse frequency derivative time series of GX 1+4 obtained from the ftp
site "gammaray.nsstc.nasa.gov/batse/pulsar/". These time series cover time span between 1991 and 1999.

\section{Timing Analysis}
\label{sect:timing}

\subsection{Pulse Timing and Pulse Frequency History}
\label{subsect:ptiming}
For the timing analysis, we follow phase coherent timing approach and use barycentric RXTE--PCA lightvurve between MJD 51974--52593. In order to search for periodicity in the data, we fold the lightcurve over trial periods (Leahy et al 1983). Using $\chi^2$ test, we obtain the template pulse profile that gives the maximum $\chi^2$. 
We create the pulse profiles with 20 phase bins for each observation and represent them in terms of Fourier harmonics (Deeter \& Boynton 1985).

\begin{figure}
\begin{tabular}{l}	\includegraphics[width=0.71\columnwidth, angle=0]{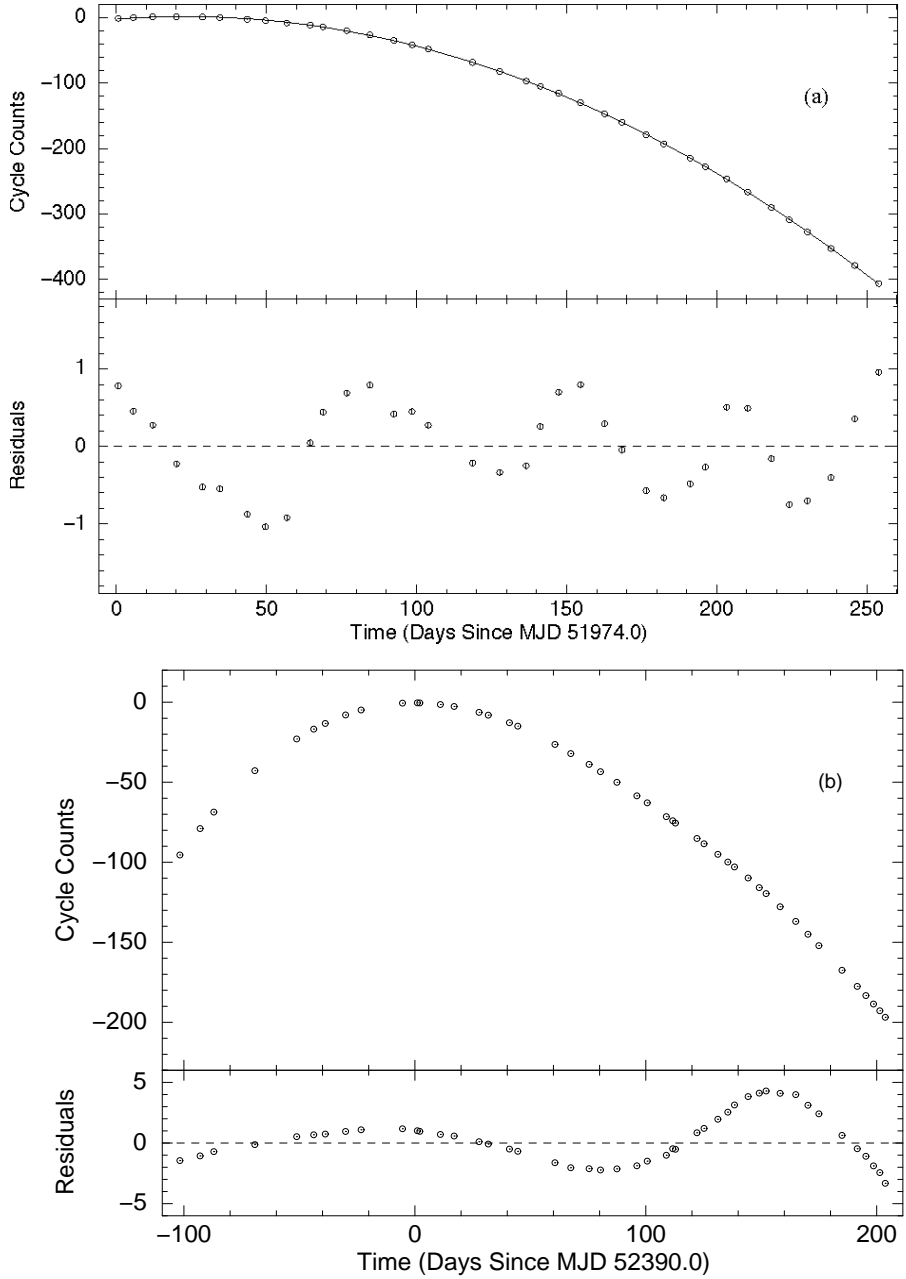} \\	\includegraphics[width=0.49\columnwidth, angle=270]{makale_phase2.eps} 
\end{tabular}
    \caption{Pulse arrival times of GX 1+4. (a) Pulse arrival times of interval "a" and its residuals after the removal of third order polynomial. 
    (b) Pulse arrival times of interval "b" and its residuals after the removal of third order polynomial.}
    \label{fig:example_phase}
\end{figure}

Then, by cross correlating each pulse with the template pulse profile, we calculate the arrival times of pulses (TOAs). In order to avoid cycle count ambiguity, we construct TOAs 
within each 50 days and measure the best period of that time span. Then, using the overlapping time intervals, we align the slopes of TOAs in consecutive time spans. And phase connected 
pulse arrival times are fitted by;
\begin{equation}
 \phi(t) = \phi_{0} + \nu_{0} (t-t_{0}) + {{1} \over {2}} \dot\nu (t-t_{0})^{2}+ {{1} \over {6}} \ddot\nu (t-t_{0})^{3} + ...
\end{equation}
where $t_{0}$ is the start time of epoch folding, $\nu$, $\dot\nu$ and $\ddot\nu$ are the spin frequency, its first derivative and second derivative, respectively. Since there is a 
gap of 60 days within the data between MJD 52228--52288, we phase connected the pulse arrival times of the source in two different time intervals "a" ($\simeq$250 days) and "b" ($\simeq$ 300 days) 
(illustrated in Figure \ref{fig:example_lc}). We are able to obtain a timing solution for each interval (parameters are listed in Table \ref{tab:timing_parameters}). We find that the source 
is spinning down with a rate of $\dot\nu\approx -2\times10^{-12}$ Hz/s. We present the pulse arrival times and their residuals after the removal of cubic trend for the intervals "a" and 
"b" in Figure \ref{fig:example_phase}.

\begin{table}
	\centering
	\caption{Timing parameters of GX 1+4 for interval "a" and "b".}
	\label{tab:timing_parameters}
	\begin{tabular}{lccr}
		\hline
		Parameter & Interval a & Interval b\\
		\hline
		MJD Range & 51974.7--52227.9 &52288.2--52593.7 \\
		Time Span (Days) & 253.2 & 305.5 \\
		Epoch (MJD) & 51974.0 & 52390.0\\
		Frequency (mHz) & 7.225369(8) &7.2571253(8)\\
		Period (s) & 136.3257(2) &137.795(6)\\
		$\dot\nu$ (10$^{-12}$ Hz/s)&-2.022(2) &-1.9892(2)\\
		$\ddot\nu$ (10$^{-19}$ Hz/s$^2$)& 0.033(2) & 1.240(4)\\
		\hline
	\end{tabular}
\end{table}

\begin{figure}
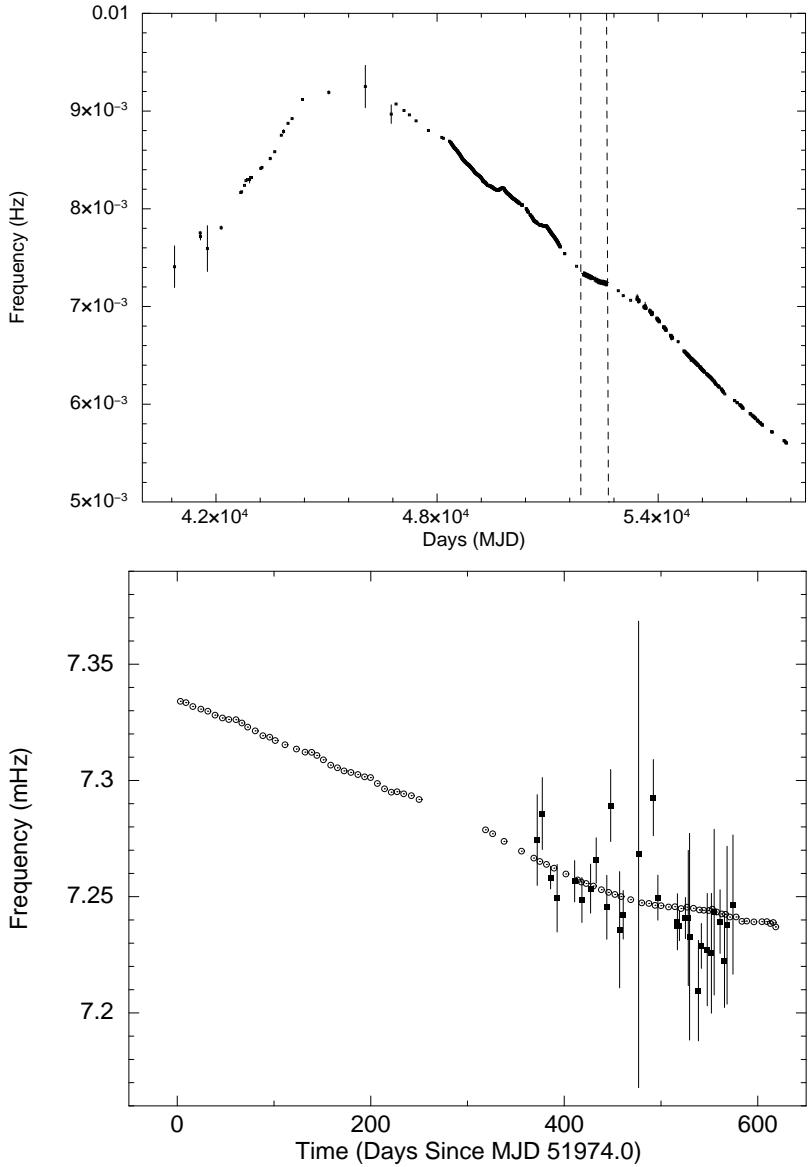

\begin{tabular}{c}
	\includegraphics[width=0.45\columnwidth, angle=270]{makale_all_freq.eps} \\
	\includegraphics[width=0.48\columnwidth, angle=270]{makale_our_freq.eps} \\
\end{tabular}
    \caption{Upper panel shows the complete frequency history of GX 1+4 (see Gonzales-Galan et. al. (2012) and references therein). 
    The frequencies measured in this work are located between the dashed lines. A closer view of this time range is given in the lower panel. Our measurements and the measurements of Cui \& Smith (2004) are represented by empty circles and filled squares respectively.}
    \label{fig:example_freq}
\end{figure}

We construct the pulse frequency history (see Figure \ref{fig:example_freq}) of the source by using the slopes of linear fits to each 3 consecutive TOAs (approximately 20-30 days). 
We calculate the error bars of frequency measurements from the uncertainties of the slopes. The pulse frequency measurements of GX 1+4 were previously conducted by 
Cui \& Smith (2004) for the time span corresponding to interval "b". Our measurements are presented together with the measurements by Cui \& Smith (2004) in the lower panel of Figure \ref{fig:example_freq}.
 
\subsection{Power Spectrum of the Pulse Frequency Derivative Fluctuations}
\label{subsect:noise}
In order to investigate torque noise characteristics of GX 1+4, we construct power spectrum of the pulse frequency derivative fluctuations by employing root mean square (rms) residuals technique developed by Cordes (1980) and Deeter (1984). In this technique,  mean square residual for the data spanning an interval of length T can be expressed as S$_{r}$T$^{2r-1}$,  where S$_{r}$
corresponds to r$^{th}$-order red noise strength.
The mean square residuals, after removing a polynomial of degree m over a time span T can be given as, 

\begin{equation} 
<\sigma _{R}^{2}(m,T)> = S_{r}T^{2r-1}<\sigma _{R}^{2}(m,1)>_{u},
\end{equation}
where $<\sigma _{R}^{2}(m,1)>_{u}$ is the normalization (proportionality
factor)  which can be estimated by measuring the variance of residuals by removing the degree of polynomial m for unit noise strength S($r=1$).

\begin{figure}
	\includegraphics[width=0.55\columnwidth, angle=270]{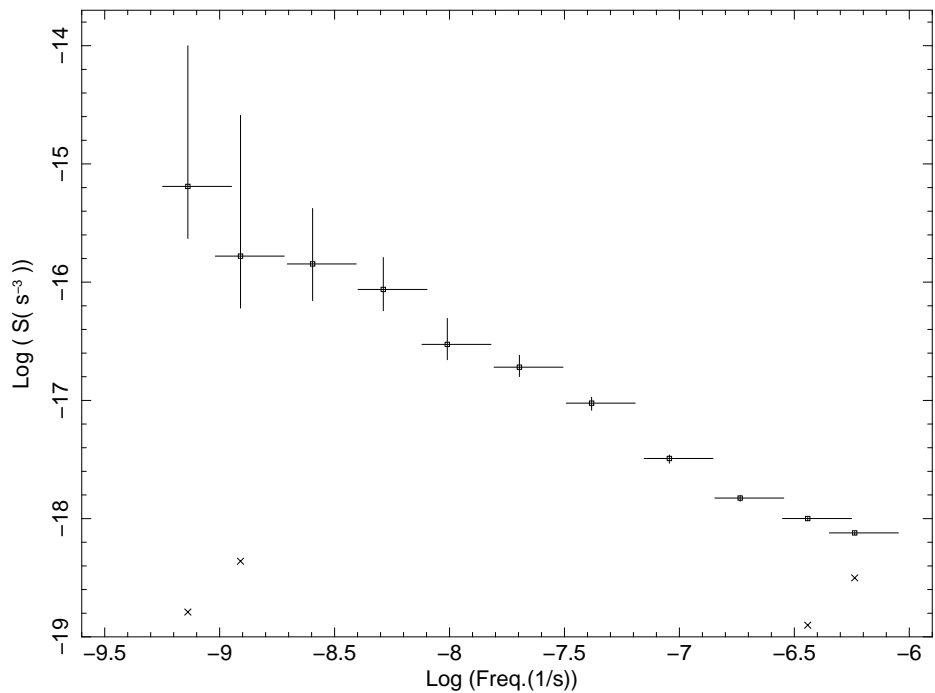}
    \caption{Power Spectrum of the pulse frequency derivatives of GX 1+4. Crosses indicate measuremental noise level.}
    \label{fig:noise}
\end{figure}

The power spectrum of pulse frequencies was constructed before from BATSE data by Bildsten et al. (1997). It was shown that  noise strengths corresponding to pulse frequency
derivatives obey $f^{-1}$ law. Therefore, we simulate time series of GX 1+4 for $f^{-1}$ noise (see Scott et al. 2003). We estimate the normalization of the simulated series 
by removing quadratic polynomial of degree $m=2$. After determining the normalization, we estimate noise strengths for different
time spans (T, T/2, T/4,...). Then, we construct power spectrum of the pulse frequency derivatives by taking the logarithmic average of the noise estimates for each time span and we present inverse of time spans as frequency in Figure \ref{fig:noise}.

Our power spectrum of the pulse frequency derivatives is constructed by using all pulse frequency measurements since 1972, whereas Bildsten et al. (1997) used only the measurements between 1991 and 1999. It is evident that the power law index and noise strengths still agree with the power spectrum estimate presented by Bildsten et al. (1997).  As seen in Figure \ref{fig:noise}, power spectrum estimate obeys $f^{-1}$ (or flicker noise) for the frequency interval from  1/44 yr$^{-1}$ 
to 1/31 d$^{-1}$. The time scale of this power spectrum is the longest ever measured for accretion powered X-ray pulsars. The noise strength level changes between $ S= 6.5 \times 10 ^{-16} $Hz s$^{-2}$  and $ S= 1.0 \times 10 ^{-18} $Hz s$^{-2}$.

\subsection{Episodic Torque Luminosity Correlations and Anti-Correlations}
\label{subsect:episodic}
Torque-luminosity correlations of GX 1+4 are examined by using CGRO-BATSE 20 -- 60 keV band X-ray flux and pulse frequency derivative time series. These time series cover a time span of $\sim 3000$ days lying within the era of long term spin-down trend of the source.  

We systematically search for X-ray flux and pulse frequency derivative correlations and anti-correlations in $\sim 100-200$ day long intervals. Assuming that the bolometric luminosity is correlated with the pulsed X-ray flux, finding such correlations or anti-correlations will be a direct indication of episodic torque luminosity correlations and anti-correlations.

\begin{figure*}
	\includegraphics[width=0.55\columnwidth, angle=90]{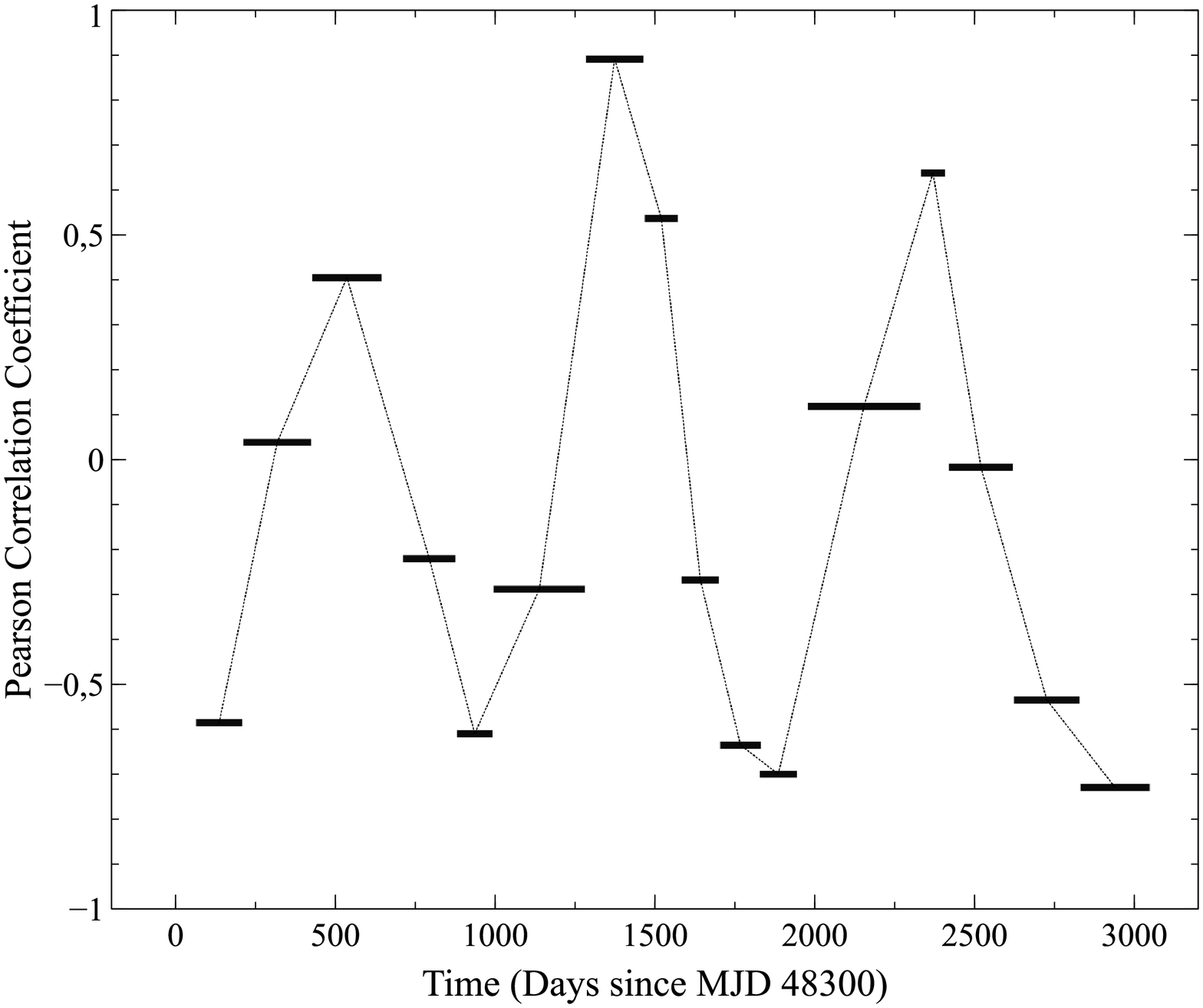}
    \caption{Time variation of Pearson correlation coefficient (PCC) between frequency derivative and pulsed flux. Correlation analysis is performed for $\sim 100-200$ day long intervals of CGRO-BATSE observations of GX 1+4.}
    \label{fig:Pearson}
\end{figure*}

We find that the source occasionally enters  $\sim 100-200$ days long intervals that show either correlation (Pearson correlation coefficient (PCC) > 0.6) or anti-correlation (PCC < -0.6) between X-ray flux and pulse frequency derivative. Outside these episodes, there is no  significant correlation or anti-correlation  between X-ray flux and pulse frequency derivative (-0.6 < PCC < 0.6). In Figure \ref{fig:Pearson}, we present the variation of Pearson correlation coefficient as a function of time for CGRO-BATSE observations of the source.   

\begin{figure}
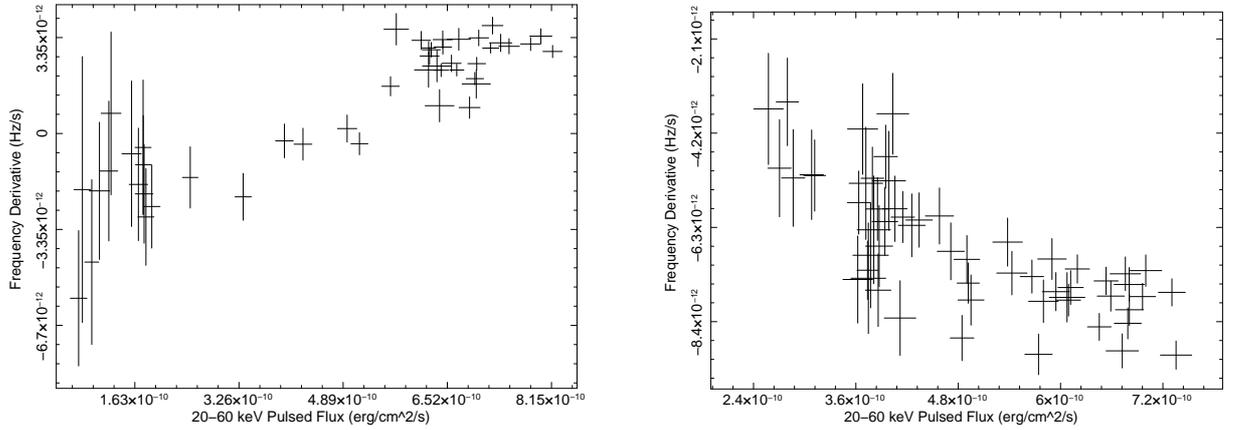

\begin{tabular}{cc}
	\includegraphics[width=0.35\columnwidth, angle=270]{positivecorr.eps} &
	\includegraphics[width=0.35\columnwidth, angle=270]{negativecorr.eps} \\
\end{tabular}
    \caption{Frequency derivative as a function of 20 -- 60 keV BATSE pulsed flux for two data sets of $\sim 180$ days long episodes that show correlation (PCC=0.89, left panel) and anti-correlation (PCC=-0.75, right panel) respectively.}
    \label{fig:corr_example}
\end{figure}

In Figure \ref{fig:corr_example}, frequency derivatives of the source are plotted as a function of 20 -- 60 keV BATSE pulsed flux for two sample ($\sim 180$ day long) intervals with mid MJD's of 49700 and 51250. These samples are the ones that show the strongest correlation and anti-correlation between frequency derivative and X-ray flux respectively. PCC for the plots on the left and right panels of Figure \ref{fig:corr_example} are 0.89 and -0.75 with the corresponding null hypothesis probabilities calculated from the Student's t-distribution (two tailed) of $1.0\times 10^{-16}$ and $4.1\times 10^{-11}$ respectively.

\section{Spectral Analysis}
\label{sect:spec}

\subsection{Time Resolved Spectra}
\label{subsect:trspectra}
Spectral fitting of individual \emph{RXTE}--PCA observations are performed in 3 -- 25 keV energy range. We basically model the spectra with an absorbed power law and 
a high energy cutoff. We also add a simple Gaussian line profile around 6.4 keV. Generally, this model successfully defines the spectra of observations with higher flux 
($\textgreater 4 \times 10^{-10}\,$erg$\,$cm$^{-2}\,$s$^{-1}$ in 3 -- 20 keV). In most of the cases, the circumstance of short ($\sim$ 3 ks) exposure does not allow to resolve the high energy 
cutoff in the spectra of lower flux observations. Even using fixed cutoff model parameters for these spectra --that have low SNR (signal-to-noise ratio)-- does not lead to better 
statistical results. Consequently, the cutoff model component is excluded while fitting 40 of 85 individual spectra. Although flux is a general criterion for model selection, there are exceptional cases. We resolve the cutoff in 7 of the low flux spectra, whereas we do not in 6 of the high flux spectra. 

\begin{figure*}
  \center{\includegraphics[width=12cm, angle=270]{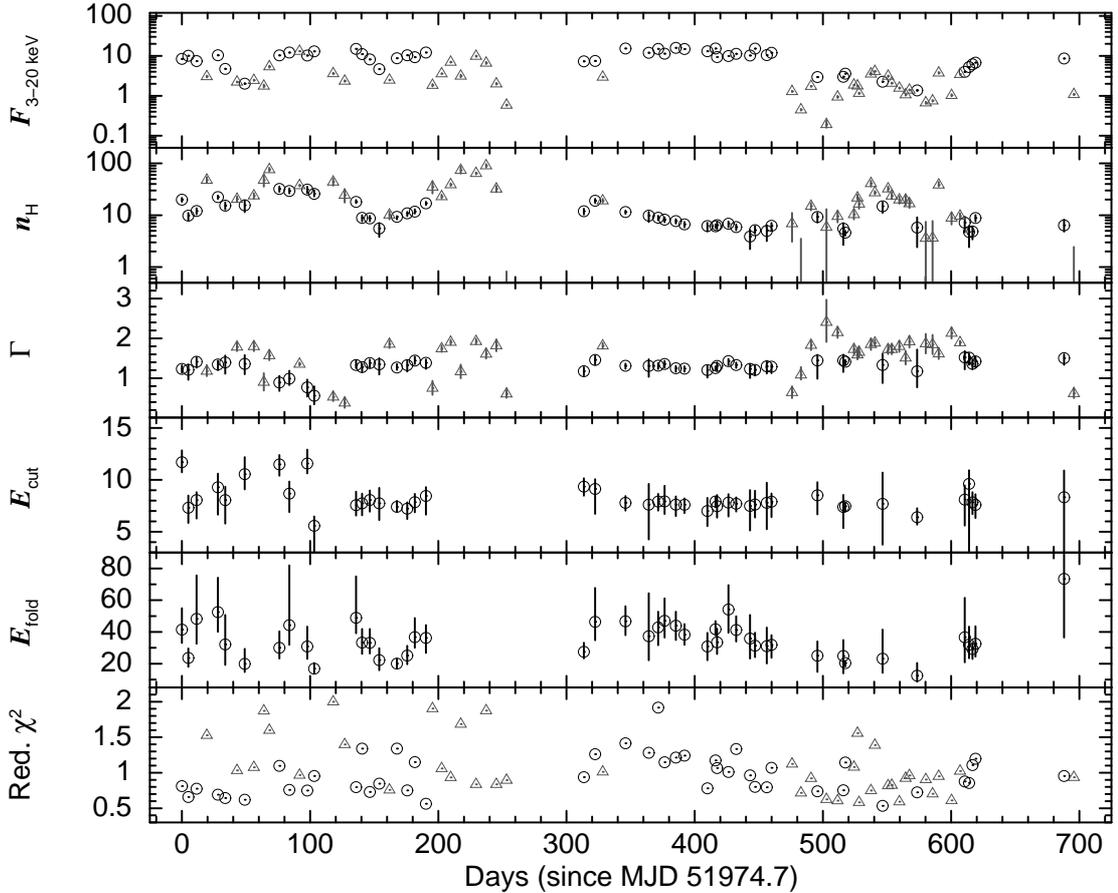}} 
  \caption{Best fit spectral parameters for individual \emph{RXTE}$-$PCA observations. Error bars indicate uncertainties at 90 per cent confidence
level. Empty circles or triangles represent parameters measured by fitting cutoff power law model [\texttt{wabs*(pow*high+gau)}] or a simple power law model [\texttt{wabs*(pow+gau)}], respectively. From top to bottom; 
variations in 3--20 keV unabsorbed flux in units of $10^{-10}\,$erg$\,$cm$^{-2}\,$s$^{-1}$, equivalent hydrogen column density in units of $10^{22}\,$cm$^{-2}$, 
photon index, cutoff energy [keV], e-folding energy [keV] and reduced $\chi^{2}$ are plotted, respectively.}
  \label{spe_par}
\end{figure*}

The time resolved changes of spectral parameters are given in Figure \ref{spe_par}. In this figure, parameters measured by fitting cutoff power law model are shown with 
empty circles, whereas parameters of a simple power law fit without cutoff are represented with empty triangles. The uppermost panel of Figure \ref{spe_par} shows the 
variation in 3 -- 20 keV unabsorbed flux. The range of flux measurements is from $0.2 \times 10^{-10}\,$erg$\,$cm$^{-2}\,$s$^{-1}$ to 
$15.9 \times 10^{-10}\,$erg$\,$cm$^{-2}\,$s$^{-1}$, hence a flux variation of two orders of magnitude is observed within $\sim$ 700 days. Equivalent hydrogen column density ($n_H$) measurements 
given in the second panel are in the range $3.6 \times 10^{22}\,$cm$^{-2}$ -- $89.9 \times 10^{22}\,$cm$^{-2}$. The $n_H$ results are demonstrated logarithmically for a better 
visualization of the variability. We commonly observe higher $n_H$ values for the spectra in which we could not resolve the cutoff. The third panel shows the variation of the photon index ($\Gamma$), which is varying from 0.4 to 2.4. Steeper power law is more common for 
the spectra in which we could not resolve the cutoff (28 of 40 no-cutoff spectra have larger indices than the average). Cutoff energy ($E_{\mathrm{cut}}$) and e-folding energy 
($E_{\mathrm{fold}}$) variations are given in fourth and fifth panels, respectively. $E_{\mathrm{cut}}$ values are consistent with a constant value of 7.8 keV. The variation of 
$E_{\mathrm{fold}}$ is stochastic; when cutoff is resolved in a low flux spectrum, $E_{\mathrm{fold}}$ is generally below 25 keV while it is usually above 30 keV for the high 
flux spectra.

A set of low flux observations are sorted through the observations that we could not resolve the cutoff individually, with the purpose of increasing the SNR of low flux spectrum. Seven observations between MJD 52498.9 -- 62642.1 (Obs.IDs: 70064-01-19-00, 70064-01-24-00, 70064-01-24-01, 70064-01-25-00, 70064-01-29-00, 70064-01-30-00, 70064-01-30-01) are combined to construct one spectrum. Here, we cautiously combine observations whose individual flux, $n_H$ and photon index values are measured to be similar. For the constructed spectrum, a simple power law fit without cutoff gives a reduced $\chi^2$ value of 2.5 which is statistically unacceptable. The fit is improved by adding the cutoff model component, resulting in a reduced $\chi^2$ value of 0.7.

\begin{figure}
  \center{\includegraphics[width=7cm, angle=270]{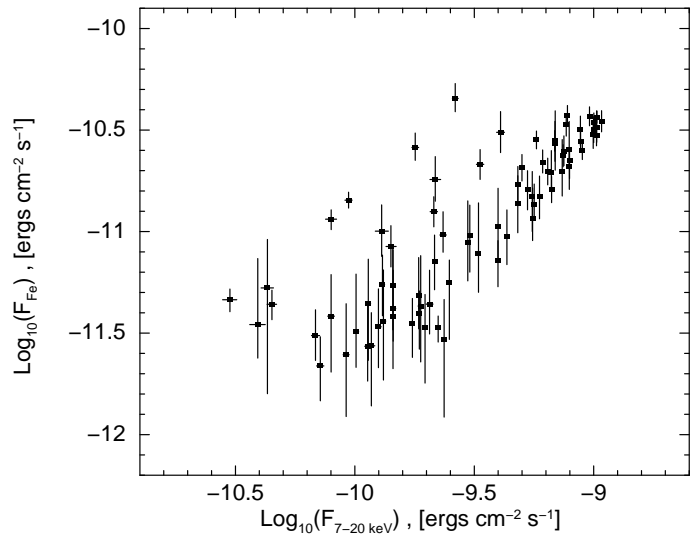}} 
  \caption{Flux variation of the \emph{Fe} emission line with respect to 7 -- 20 keV unabsorbed flux from \emph{RXTE}$-$PCA observations. Single parameter errors are 
  calculated at 90 per cent confidence level.}
  \label{Fe}
\end{figure}

We also check whether the time variability of spectral parameters is related to X-ray flux or not. We only find a significant correlation for \emph{Fe} line flux. For individual \emph{RXTE}--PCA 
spectra, the flux contributions of the \emph{Fe} line are measured using the \verb"CFLUX" model in \verb"XSPEC" and they are plotted against 7 -- 20 keV flux measurements in Figure \ref{Fe}. 
In order to test the significance of the correlation, we calculate the Pearson correlation coefficient between the two parameters and it is found as 0.82. The null-hypothesis 
probability calculated from the Student's t-distribution (two-tailed) is $2.3 \times 10^{-20}$.

\subsection{Pulse Phase Resolved Spectra}
\label{subsect:pprspectra}
\begin{figure*}
  \center{\includegraphics[width=13.8cm, angle=270]{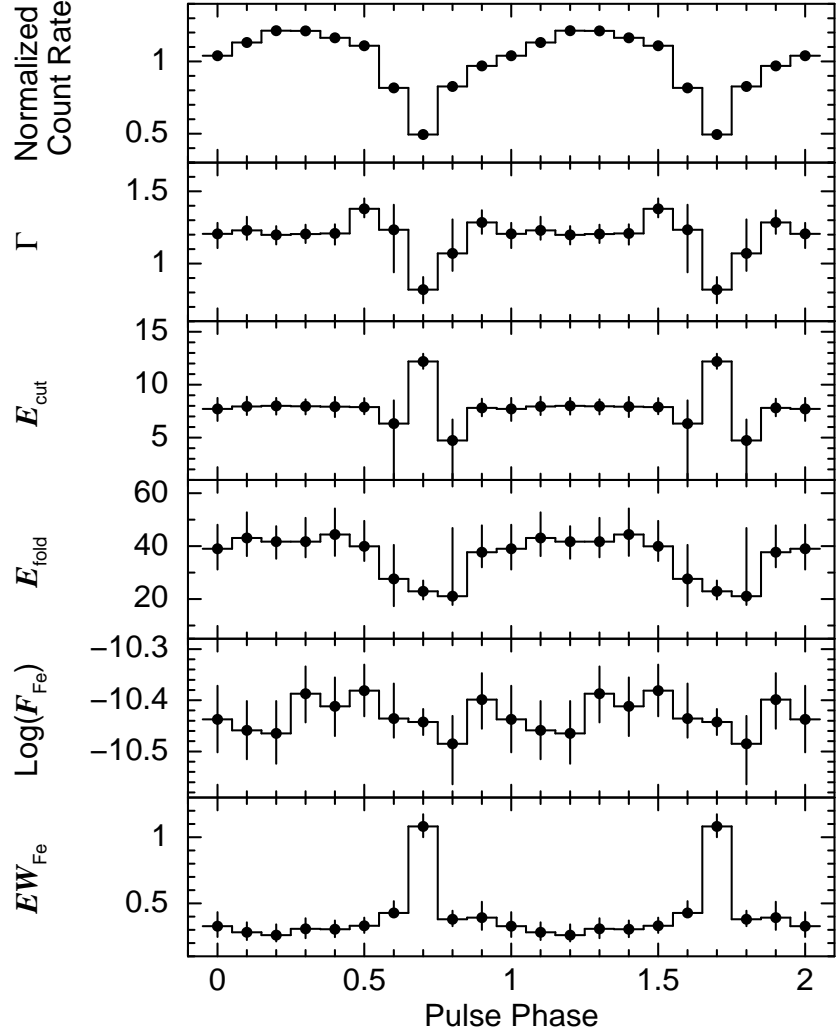}} 
  \caption{Best fit spectral parameters of the 10 phase-binned spectra from the \emph{RXTE}$-$PCA observation sequence on MJD 52390.9.
  The pulse profile is given at the uppermost panel for comparison.
  All uncertainties are calculated at the 90 per cent confidence level. For clarity, the data points are repeated for a cycle.}
  \label{fase_high}
\end{figure*}

We construct pulse phase resolved spectra of GX 1+4 from the observation sequence on MJD 52390.9, which is one of the brightest observations with a long exposure time. 
The timing solution for the corresponding time is appended to the timing file of the \verb"FASEBIN" tool and the pulse period is divided into ten equal segments each of which 
having an exposure of $\sim$ 1.8 ks. The cutoff power law  model is used for fitting of phase-binned spectra in 3--25 keV. The results are plotted in 
Figure \ref{fase_high}.

The spectral parameters that change with pulse phase are photon index, cutoff and e-folding energies. The spectral bin that corresponds to the pulse minimum has significantly higher cutoff energy, lower photon index and lower e-folding energy. This indicates a harder spectrum for this phase. For this model; equivalent hydrogen column density does not change with pulse phase, it is consistent with a constant value of $6.0 \times 10^{22}\,$cm$^{-2}$.

For the Gaussian emission representing the iron K$\alpha$ line around 6.4 keV, the equivalent width is the only parameter of the line that varies with pulse phase. The spectrum of the phase that coincides with the pulse minimum shows an increase in the EW of Fe line, without any significant change in line flux. 

\section{Discussion}
\label{sect:discuss}

\subsection{Timing Solution and Frequency History}
GX 1+4 is a persistent accretion powered X-ray pulsar and a peculiar source residing in a symbiotic X-ray binary. Long term spin rate evolution of the source has been monitored continuously since it was discovered in 1970s (see Gonzales-Galan et al. (2012) and references therein). From the pulse timing analysis of RXTE-PCA observations, we are able to phase connect the pulse arrival times of the source within two different time intervals of  $\sim 250$ and $\sim 300$ days long as shown in Figure \ref{fig:example_phase}. Therefore, we obtain timing solutions  corresponding to these two intervals (see Table \ref{tab:timing_parameters} including interval-wise measurements of the first and second derivatives of pulse frequency. Morever, using these pulse arrival times, we contributed to long term pulse frequency history of the source with our new pulse frequency measurements (see Figure \ref{fig:example_freq}).     

GX 1+4 was a spinning-up source in 1970s before it underwent a torque reversal in 1980s. The source was found to show correlation between spin-up rate and X-ray flux before the torque reversal, which was interpreted as an indication of a persistent prograde accretion disc (see Ghosh\& Lamb 1979, Wang 1987).

After the source underwent torque reversal, the prograde accretion disc scenario was found to be inconsistent with the spin rate and X-ray flux behaviour of the source. By using CGRO-BATSE data, Chakrabarty et al. (1997b) found that there is a general anti-correlation trend between spin rate
and pulsed flux (or in other words correlation between negative of spin rate and pulsed flux) during the continuous spin-down interval between 1991 and 1995, which is the opposite of expected in the presence a prograde accretion disc. However, they also reported that anti-correlation state of GX 1+4 is not
perpetual and there is a marginal evidence of a positive correlation between X-ray flux and spin rate for a prolonged spin-up ($\sim$ 200 days long) interval. Similar spin rate and X-ray flux anti-correlation was also found
from the analysis of $\sim$ 600 day long Fermi/GBM and Swift/BAT data (Gonzales-Golan et al. 2012). In order to investigate the correlation state from RXTE-PCA observations, we measure spin down rates for 3 different time intervals with different flux levels (see Table 3).
The measurements indicate that spin down rate is scaling up with the flux level, therefore a general anti-correlation state can be inferred during RXTE observations.

\begin{table}
\label{fig:flxlvl}
 \centering
 \caption{RXTE--PCA spin rate measurements for the intervals with different flux levels}
 \begin{tabular}{l|c|c|c}
  & Interval 1 & Interval 2 & Interval 3 \\
 \hline
  MJD Range & 51974.7-52227.9 & 52288.2-52343.6 & 52450.6-52593.7\\
  Time span (days) & 253.2 & 146.4 & 143.1 \\
  $\dot{\nu}$ ($10^{-12}$ Hz/s) & -2.0585(3) & -2.3860(7) & -0.8334(8) \\
  3--20 keV Unabsorbed Flux ($10^{-10}$ erg cm$^{-2}$ s$^{-1}$)& 6.22(2) & 9.90(3) & 2.21(1) \\
\hline
 \end{tabular}
\end{table}

\subsection{Torque Noise Strength}

Torque noise fluctuations and noise strengths of pulse frequency fluctuations have been studied for several 
accretion powered X-ray pulsars  (Bildsten et al. 1997; Baykal \& \"{O}gelman 1993).  Red noise (random walk) in pulse frequency fluctuations or white noise in the pulse frequency derivatives are resonable models for wind accreting X-ray binaries like Vela X-1, 4U 1538-52 and GX 301-2 (Bildsten et
al., 1997). These sources have power spectra with white noise strengths in the range $10^{-20}-10^{-18}$ Hz s$^{-2}$. The persistent long-term spinning-down source 4U 1907+09 (Baykal et al. 2001, Baykal 2006) also shows random walk in the pulse frequency history (\c{S}ahiner
et al. 2012) with noise strength $1.27 \times  10^{-21}$  Hz s$^{-2}$ for a frequency interval between 1/1300 d$^{-1}$ and 1/75 d$^{-1}$. For this source, formation of an episodic transient accretion disc around the neutron star was suggested to explain the random walk model in pulse frequency (\c{S}ahiner
et al. 2012). 

Accretion powered pulsars in low mass X-ray binaries like Her X-1 and 4U 1626-67 accrete via persistent accretion discs and their pulse frequency time series are also consistent with the random walk model. Their pulse frequency derivatives have white noise strengths in the range $10^{-21}$ to $10^{-18}$ Hz s$^{-2}$. Since the power spectra of these sources lie in a narrow range, possibility of red noise in the pulse frequency derivative can not be excluded (Bildsten et al. 1997).

Cen X-3 accretes via disc and and the noise strength varies from low to high frequencies as $10^{-16}$, $10^{-18}$ Hz s$^{-2}$ (Bildsten et al. 1997).The power law index of the power spectrum in this
system is $\sim-1$ which implies that disc accretion dominates at short time scales.

OAO 1657-415 has a power law index of $\lesssim 0$ with noise strength $10^{-17}$ Hz s$^{-2}$ (Baykal 1997). This source also shows marginal correlation with spin-down rate and X-ray luminosity (Baykal 2000). 

X Per has the lowest noise strength among accretion powered pulsars in high mass X-ray binaries with $10^{-21}-10^{-23}$ Hz s$^{-2}$ for frequencies between 1yr$^{-1}$ and 1/35 yr$^{-1}$ (Acuner et al. 2014). The steep power law index in pulse frequency derivative indicates that it could have a transient accretion disc.

SAX J2103.5+4545 has the  steepest power law index with 2.13 
among high mass X-ray binaries (Baykal et al. 2007). The steep power law index suggests that accretion via accretion disc
at shorter time scales possesses low timing noise therefore
 power density spectrum becomes more steeper compared to the other persistent high mass X-ray binaries. 

In GX 1+4, we observe flickering noise ($f^{-1}$) which agrees with the power spectrum estimate of Bildsten et al. (1997). It is important to note that we extend the estimate for a time scale ranging from 31 days to 44 years. Thus, we achieve to describe a noise process for the longest timescale among accretion powered X-ray pulsars. We find that the noise strength level of the source changes between $ S= 6.5 \times 10 ^{-16} $Hz s$^{-2}$  and $ S= 1.0 \times 10 ^{-18} $Hz s$^{-2}$.

The  torque noise power spectrum of GX 1+4 has a unique trend among accretion powered pulsars, 
showing $f^{-1}$ power law trend in the longest timespan. In other words, timing noise increases with time. Recently Ilkiewicz et al. (2017) supported the orbital period of 1162 days using hard (>17 keV) X-ray observations. This orbital period was originally suggested by Hinkle et al. (2006) using radial velocity measurements from infrared observations. 
 Hinkle et al. (2006) found the projected semi-major axis of the orbit as  $a/c \sin i \sim 773$ lt s. The effect of Doopler shifts due to the orbital modulation on pulse frequency time series is thus
$\delta \nu \sim {{2 \pi} \over {P_{orbit}}}{{1} \over {P_{spin}}} {{a} \over {c}} \sin i \sim 3.7 \times 10 ^{-7}$ Hz. Using the torque noise spectrum (see Figure \ref{fig:noise}), the noise strength at the frequency corresponding to orbital period ($1/(1162*86400)= 0.9 \times 10 ^{-9}$ Hz) is $S \sim 3 \times 10 ^{-17} Hz s^{-2}$, we estimate the
deviation on frequencies due to the noise process as 
$\sqrt{\Delta \nu^2} = \sqrt{ST} = 5.5 \times 10^{-5}$ which is two orders of
magnitude greater then Doppler shifts expected due to orbital modulation. 
Therefore it is not possible the resolve Doppler shift modulations in pulse
frequency time series, since the noise process dominated. Moreover, $\sim 700$ days observation time of RXTE is short compared to the suggested orbital period. Continuous monitoring of the source using future observatories such as LOFT (Ferroci et al . 2012) might be useful to resolve Doppler shift modulations from pulse arrival data and obtain orbital parameters of the system from X-ray observations.

\subsection{Episodic Correlations and Anti-correlations between X-ray flux and Spin Rate}

In contrast to previous studies that concentrates on searching for "general" correlation or anti-correlation states between X-ray flux and spin rate (Chakrabarty et al. 1997b; Paul et al. 1997; Gonzales-Golan et al. 2012), we search for episodic correlations and anti-correlations on shorter time scales ($\sim$100-200 day long) by using CGRO-BATSE measurements. We find that the correlation state of GX 1+4 seems to change on $\sim$ 100-200 days long intervals (see Figure
\ref{fig:Pearson}). Furthermore, the correlation and anti-correlation intervals do not show significant X-ray flux changes compared to uncorrelated intervals. Still, we might argue that the correlation and
anti-correlation intervals might be indications of transient accretion disc formation (either prograde or retrograde) and episodic variation of accretion geometry without significant X-ray flux
variations.

From Figure \ref{fig:Pearson}, we also note that variation of Pearson correlation coefficient (PCC) has a period of about 1000 days. Flicker type noise
process suggests that the torque events are correlated or at least carries a
memory from past history. This periodicity of the correlation states is consistent with the proposed orbital period of 1162 days (Hinkle et al. 2006; Ilkiewicz et al. 2017) suggesting that the alteration of the states of positive and negative correlation states continues for several decades of orbital cycles. The accretion process in GX 1+4 is unique as it is qualitatively seen from correlation -- anti-correlation episodes and torque noise power law trend.    

Future monitoring observations of GX 1+4 will be useful to further understand the
torque $-$ X-ray luminosity relation.

\subsection{Energy Spectrum}

We investigate X-ray energy spectrum of the source using absorbed power law model with or without cut-off, including a Gaussian line for Iron line complex. Analysis of the time resolved spectra by using individual RXTE-PCA observations indicate a significant correlation between iron line flux and unabsorbed X-ray flux for the first time for GX 1+4. Similar correlation was found before for the accretion powered X-ray pulsars OAO 1657-415 (Jaisawal \& Naik 2014), SWIFT J1729.9--3437 (\c{S}ahiner et al. 2013), Her X-1 (Naik \& Paul 2003; \.{I}nam \& Baykal 2005) and LMC X-4 (Naik \& Paul 2003).  This correlation might be an indication of the fact that the iron lines originates from the cool matter near the neutron star (Makishima 1986).

Pulse phase resolved spectra of the source are also studied. Using the same spectral model used for the analysis of time resolved spectra, it is found that the spectral bin corresponding to the pulse minimum has significantly higher cutoff energy, lower photon index and lower e-folding energy indicating a harder spectrum for pulse minimum. Moreover, the same pulse phase shows an increase in equivalent width of Fe line whithout any significant change in line flux. A constant line flux over all the pulse phases might be an indication of the fact that the line emitting region surrounds the pulsar, whereas broadening of Fe line on continuum of minimum phase might be a consequence of the blending of K$\alpha$ and K$\beta$ lines at 6.4 and 7.05 keV, while K$\beta$ at 7.05 keV is smothered due to enhanced emission on the continuum of other phases.

\section*{Acknowledgment}

We acknowledge support from T\"{U}B\.{I}TAK, the Scientific and Technological Research Council of Turkey through the 
research project MFAG 114F345. We would like to thank the anonymous referee for valuable comments which helped to improve the manuscript.






\bsp

\label{lastpage}

\end{document}